\documentclass[12pt,letter]{article}

\usepackage[latin1]{inputenc}
\usepackage{graphicx,float,xspace}

\newcommand{\Iac}{\ensuremath{I_{AC}}\xspace}
\newcommand{\Ic}{\ensuremath{I_{C}}\xspace}
\newcommand{\Imm}{\ensuremath{I_m}\xspace}
\newcommand{\Idc}{\ensuremath{I_{DC}}\xspace}

\newcommand{\Isb}{\ensuremath{I_{sb}}\xspace}

\begin{document}
\newpage

\title{\vspace{10pt}\flushleft 
\textbf {\sc Decrease of ac losses in high Tc superconducting tapes 
by\\
application of a dc current
}
}
\date{\em Proceedings of CEC-ICMC july 12-16, 99 (Montréal) -- talk 
IBB6}
\author{
{B. des Ligneris$^1$, Marcel Aubin$^1$, J. Cave$^2$, W. Zhu$^2$, P. 
Dolez$^3$.\vspace{12pt}}
\\
{$^1$ \footnotesize Département de physique and Centre de recherche en 
physique du solide}
\\
{ \footnotesize Université de Sherbrooke, Sherbrooke, Québec, Canada 
J1K 2R1}
\\
{ $^2$ \footnotesize Technologies Émergéntes de production et de 
stockage, VPTI Hydro-Québec}
\\
{ \footnotesize Québec, Canada J3X 1S1}
\\
{ $^3$ \footnotesize Virgina Tech University -- Virginia -- USA}
}
\maketitle

\thispagestyle{empty}

\section*{\sc Abstract}

The ac losses in a silver-gold alloy sheathed Bi-2223 tape were 
measured by the null calorimetric method while a dc current was 
superimposed onto the ac current.  As expected through computer 
simulations, a Clem valley was observed for the larger ac currents.  A 
reduction of ac losses of approximately 50\% at the valley minimum is 
observed in accord with the calculations.  At lower ac currents, the 
data fit the calculated behavior with a single parameter, a Meissner 
current of 21.8~A.  It is shown that a surface barrier current is 
incompatible with the data.

\section*{\sc Introduction}

The question of ac losses in superconducting tapes was raised long 
before the advent of high Tc superconductors.  For example, McConnell 
and Critchlow \cite{connel75} reported ac loss measurements on NbTi 
wires in 1975.  On the theoretical side, Bean \cite{bean64} proposed 
his now well known critical state model in 1964 while considering the 
hysteresis losses in a type II superconductor due to alternating 
magnetic fields.  According to this model, the losses increase as the 
third power of the magnetic field amplitude (for amplitudes that are 
small with respect to the full penetration field).  A similar behavior 
was predicted by Norris \cite{norris70} when a transport current 
replaces the magnetic field as long as a circular or elliptical cross 
section is considered.  In the case of a thin strip however, the losses 
vary as the fourth power of the current, again for small currents.  
These behaviors have all been confirmed experimentally by a number of 
workers\cite{pat96}.  

The fact that ac losses always seem to increase with current or 
magnetic field amplitude represents a major concern in the pursuit of 
superconductivity applications.  An interesting exception occurs 
however when a dc magnetic field or current is superimposed onto the ac 
field or current.  It was observed by Thompson, Maley and Clem5 that an 
increasing dc-bias magnetic field added to an ac field decreases the 
losses initially before the inevitable increase occurs. The minimum 
defines what is now known as the Clem valley.  It was only recently 
that this valley was observed with transport currents \cite{ben98} 
rather than magnetic fields.  Furthermore, this observation involved 
high Tc superconducting tapes, more specifically those of silver-gold 
alloy sheathed Bi--2223.  While these were the first measurements 
showing explicitely the Clem valley in high Tc materials, a reduction 
in losses can be seen in the data of  Oomen et al\cite{oomen97} in 
their plot of losses as a function of magnetic field amplitude with and 
without a dc current as they attempted to separate intragrain and 
intergrain effects.  Their curve obtained with a dc current is lower 
than the one without a dc current.

In this paper, we extend the measurements and analysis of the data 
illlustrating the Clem valley with transport dc and ac currents in 
Bi--2223 silver-gold alloy sheathed tapes.  More data points have been 
obtained and the computer simulations include surface barrier effects 
as well as Meissner currents and take into account the elliptical 
cross-section of the tape.

\begin{figure}[h]
\centering
\includegraphics[width=5in]{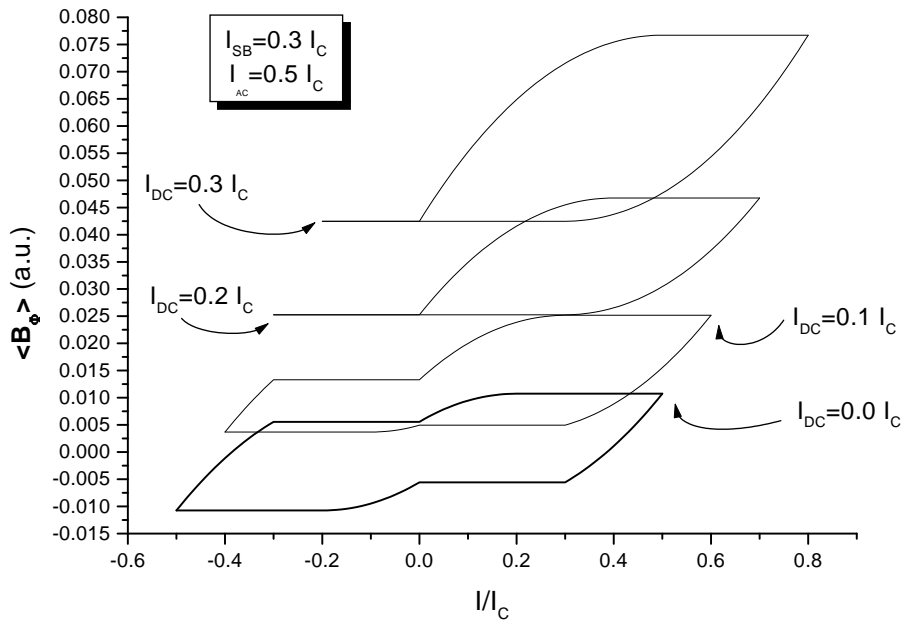}
\caption{\footnotesize Hysteresis loops with fixed surface barrier and 
alternative currents.}
\label{hyster_isb_vallee}
\end{figure}

\section*{\sc Theoretical considerations}

We will proceed with the Bean model which, in spite of its simplicity 
has had remarkable success in interpreting data on ac losses in 
superconductors following the application of ac magnetic fields or ac 
transport currents.  In our case however, complications arise due to 
the addition of a dc current to the ac current.  The flux density 
profiles are no longer symmetric.  Furthermore we include the 
possibility of either a  Meissner current \Imm or a surface barrier 
current \Isb.  Finally we do not restrict ourselves to a tape of 
circular cross-section but consider one with an elliptical 
cross-section in the computer simulations of the ac losses.  LeBlanc et 
al\cite{leblanc96} had performed such simulations to show in detail how 
the reduction of ac losses comes about with the addition of a dc 
current in the context of cylindrical symmetry.  We follow the lead of 
LeBlanc et al except for consideration of an elliptical cross-section 
and extend the project to fit experimental curves of the phenomenon.  
Our consideration of an elliptical cross-section inserts a form factor 
into the analysis given by

\begin{equation}
\int_{0}^{2\pi}{\sqrt{1-e^2.cos(p^2)}dp}
\end{equation}

where e is the eccentricity of the ellipse.  More details on the 
simulation calculations will be given elsewhere\cite{ben99}.

\begin{figure}[h]
\centering
\includegraphics[width=5in]{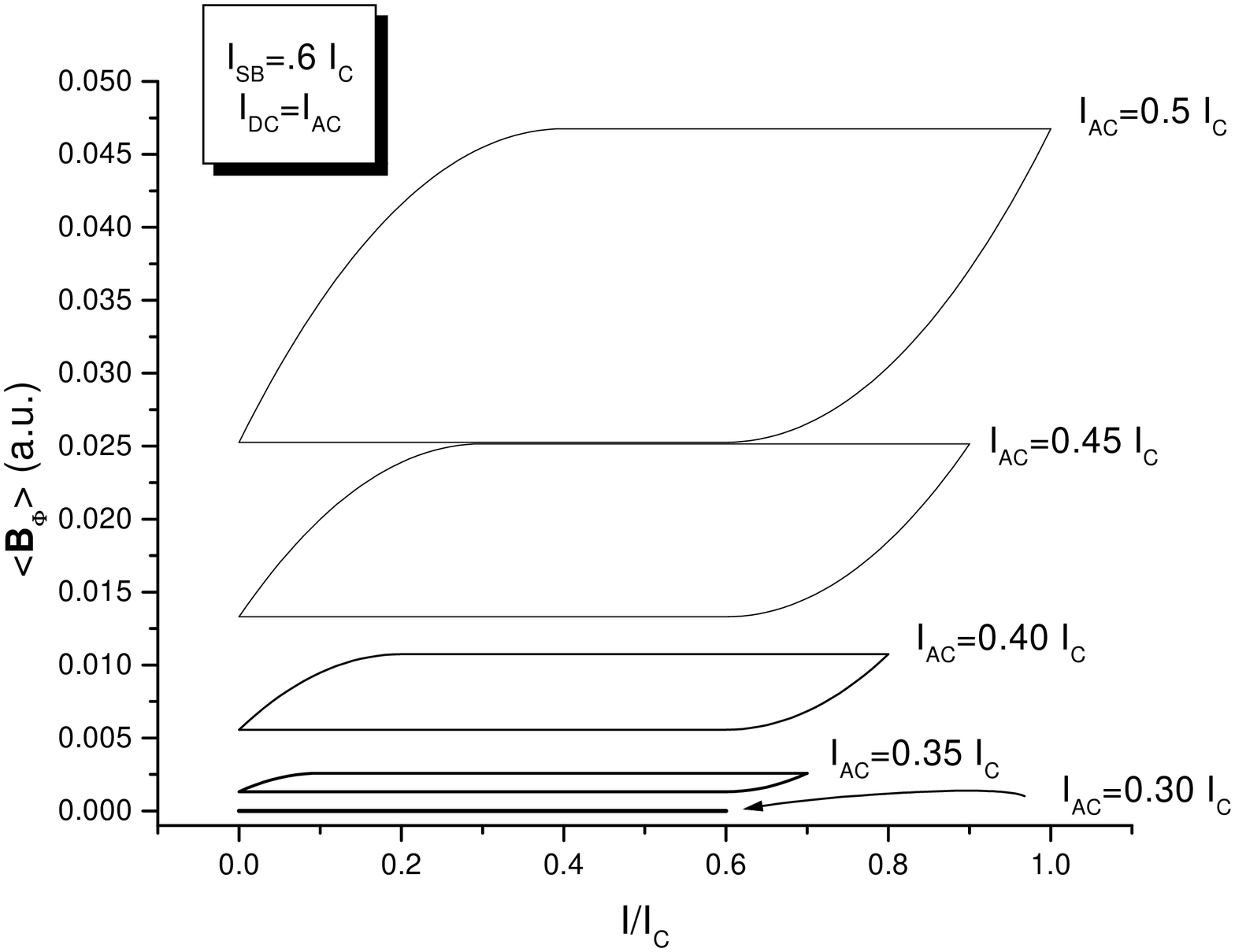}
\caption{\footnotesize Hysteretis loops with a fixed surface barrier 
current, \Idc~=~\Iac and \Iac~$\le$~\Isb.}
\label{hyster_isb_plateau}
\end{figure}

We show in Figures \ref{hyster_isb_vallee} and 
\ref{hyster_isb_plateau}, the spatial average of the magnetic flux 
density as the ac current goes through one cycle.  The area enclosed by 
a hysteresis loop is proportional to the ac loss per cycle.  Each loop 
has four horizontal lines of length \Isb.  Note that all currents are 
expressed in terms of the critical current \Ic.  In Figure 
\ref{hyster_isb_vallee}, Iac is larger than Isb.  For zero dc current, 
the loop is symmetric with respect to the abscissa.  For increasing dc 
currents, the loops shift upwards and to the right and the horizontal 
lines of the left side of each loop approach each other and eventually 
merge when \Idc~=~\Iac~-~\Isb such that the net area of the loop 
decreases.  It is this decrease that leads to the decrease in ac losses 
and to the Clem valley.  

In Figure \ref{hyster_isb_plateau}, ac currents less than Isb are 
considered.  We fix \Idc~=~\Iac such that the maximum losses for the 
given ac current are calculated.  With this representation the minimum 
current for each loop corresponds to zero total current in order to 
illustrate an important result.  The loops decrease in area with 
decreasing ac current until the zero area condition is reached i.e. 
\Iac equal to or less than \Isb/2.  Another important result can be 
found if \Idc~$\leq$~\Iac a condition found in this figure.  If one 
were to increase Idc while maintaining \Iac constant, the loop would 
merely shift upwards and to the right without changing in area.  Indeed 
this is the maximum area for the given ac current as implied above.

\begin{figure}[h]
\centering
\includegraphics[width=5in]{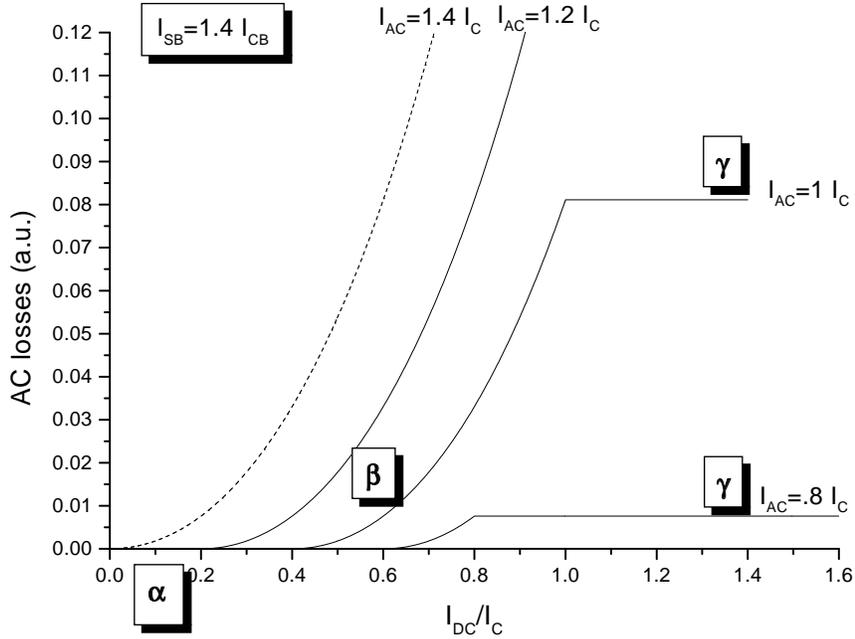}
\caption{\footnotesize AC losses with a fixed surface barrier current 
and an AC current greater than half the surface barrier current. Regime 
$\alpha$ occurs when no losses arise, regime $\beta$ begins when losses 
increase and regime $\gamma$ occurs when losses are saturated 
(plateau).}
\label{pertes_isb_plateau}
\end{figure}

The evolution of the ac losses with dc current is summarized in Figure 
\ref{pertes_isb_plateau} although \Isb has been increased to encompass 
a larger number of situations.  When \Iac~=~\Isb, losses appear for any 
non-zero dc current.  For smaller values of \Iac, no losses are 
encountered below a threshold dc current (the $\alpha$ region defined 
in Ref.\cite{leblanc96}).  If one reduces the ac current even further 
such that \Iac~$\le$~\Isb/2 no losses are encountered regardless of the 
dc current. Beyond the $\alpha$ region, losses appear as in the $\beta$ 
region. The $\gamma$ region is the plateau which appears suddenly as 
soon as the dc current reaches \Iac.  This corresponds to the loop area 
becoming constant as discussed in the context of Figure 
\ref{hyster_isb_plateau}.

\begin{figure}[h]
\centering
\includegraphics[width=5in]{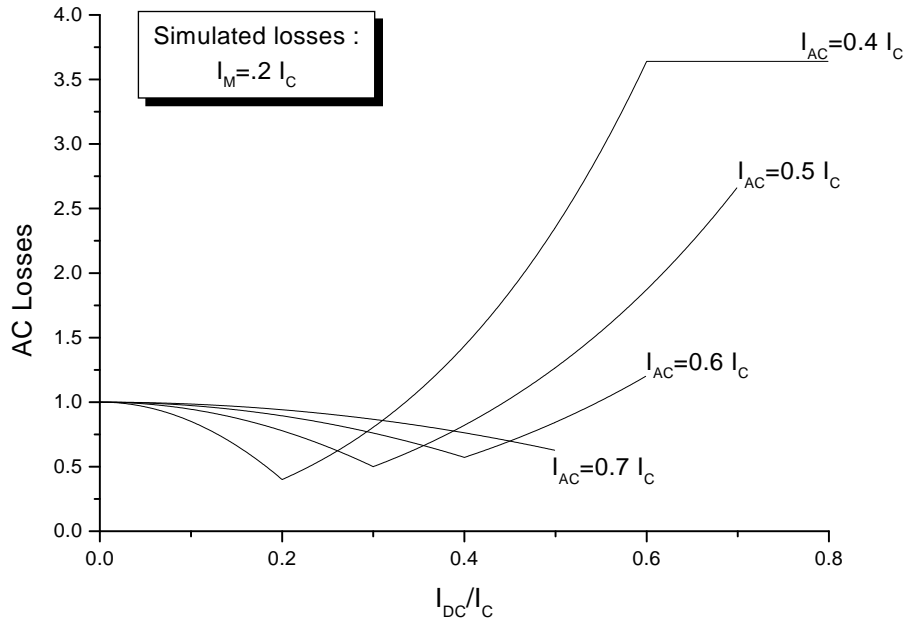}
\caption{\footnotesize Clem valley~: AC losses with a meissner current 
and \Iac~$>$~\Imm}
\label{pertes_im_vallee}
\end{figure}

The first three figures have described the situation with a surface 
barrier current \Isb.  With a Meissner current instead,  there is a 
horizontal line of length 2~\Imm in the center of the upper part of the 
loop and a similar one in the lower part of the loop as illustrated in 
Ref.\cite{ben98,leblanc96} (rather than four horizontal lines of length 
Isb at the extremities of the loop as in Figure 1).  A consequence of 
this is the absence of a  zero loss condition analogous to the one with 
a surface barrier current (\Iac less than \Isb/2) i.e. losses arise 
when \Idc~+~\Iac~$<$~\Imm, a condition that can always be fulfilled by 
increasing \Idc.  The general appearance of the $\alpha$,$\beta$ and 
$\gamma$ regions remains although with different conditions.  As with 
\Isb, \Imm leads to a Clem valley in the losses for large values of 
\Iac as shown in Figure \ref{pertes_im_vallee}. Here, the calculated 
losses are normalized with respect to the zero bias condition. These 
behaviors are similar to the ones predicted by LeBlanc et al. who 
considered a cylindrical geometry.

\section*{\sc Results and discussion}

The measurements reported here were obtained on a silver-gold alloy 
sheathed Bi--2223 tape fabricated by the powder-in-tube method in the 
Hydro-Québec laboratories.  It is monofilamentary with a cross-section 
of 2.10$^{-3}$~cm$^2$ and a length of approximately 30~cm.  The 
critical current was determined by fitting the V -- I characteristic 
with the double integration of the sum of gaussian curves.  Initially a 
value of 29~A was obtained but excessive heating decreased this value 
to 25~A. Thus data will be presented with both values of critical 
current. The losses were determined by the null calorimetric technique 
which is described elsewhere \cite{pat96,pat98_2}.  This technique was 
developed by the authors over the last few years and has the advantage 
of measuring the total losses as well as being adaptable to complex 
environments.  

\begin{figure}[h]
\centering
\includegraphics[width=5in]{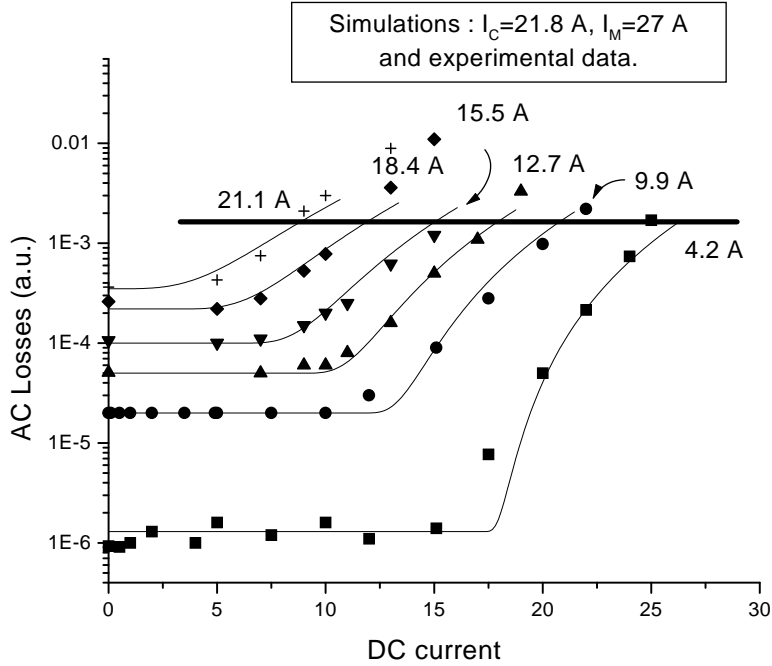}
\caption{\footnotesize Measured AC losses as a function of DC current. 
The curves are the results of our simulations  with a Meissner current 
(parameters~: \Ic=~27~A and \Imm=~21.8~A for all curves).}
\label{pertes_exp_plateau}
\end{figure}

We show the results with a critical current of 29~A and low ac currents 
in Figure \ref{pertes_exp_plateau}.  These data were taken at 559 Hz 
rather than at a lower frequency to increase the losses and thus the 
signal to noise ratio since the losses are low when the ac current is 
less than half the critical current.  It had been verified that the 
frequency does not affect the shape of the curves.  The \Iac values in 
the figure correspond to the amplitude of the current as defined above 
and not to the rms readings of the instrumentation.  The curves through 
the points are the results of a fit to be discussed below.

\begin{figure}[h]
\centering
\includegraphics[width=5in]{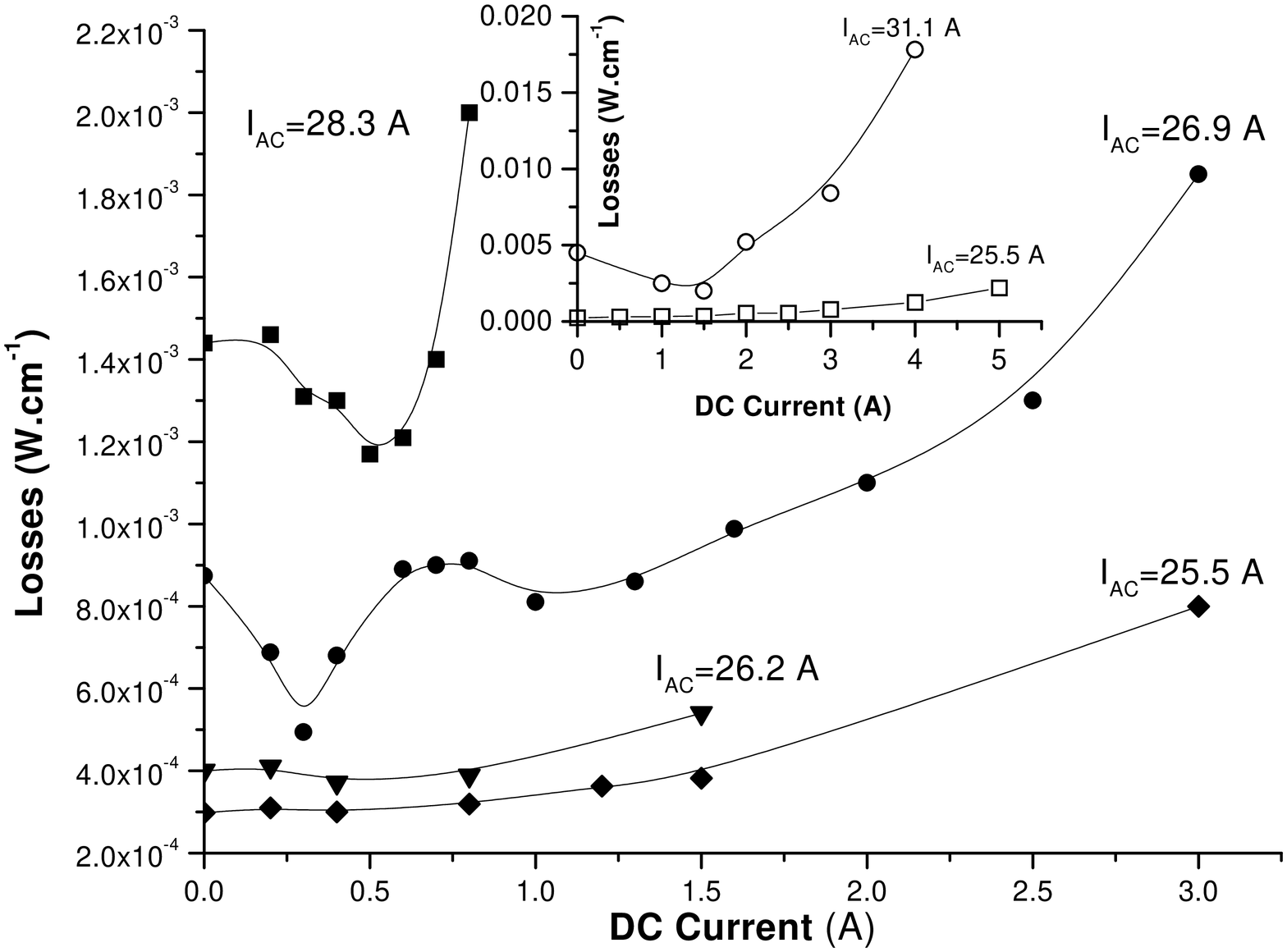}
\caption{\footnotesize Measured AC losses~: Clem valley. The main 
pannel is with \Ic=~25~A and the inset with \Ic~=~29~A.}
\label{pertes_exp_vallee}
\end{figure}

At higher ac currents the data were taken at the lower frequency of 55 
Hz since the losses are greater and more easily measured.   Again with 
a critical current of 29~A, a minimum in the losses is observed (inset 
of Figure \ref{pertes_exp_vallee}) as anticipated in the theoretical 
section above.  After the critical current had been reduced to 25~A, 
even clearer minima were observed as shown in the main part of Figure 
\ref{pertes_exp_vallee}.

Returning now to the low ac current data of Figure 
\ref{pertes_exp_plateau}, we note that the losses display an initial 
plateau for each ac current value before the increase, reminiscent of 
the a region of Figure \ref{pertes_isb_plateau}.  The non-zero level of 
the plateau however is perplexing, unless we recall the granular nature 
of the Bi--2223 tape.  The intergrain material has a low critical 
current density and act as weak links between the grains.  Even in the 
absence of a dc current and with low ac currents, vortices are free to 
move in the intergrain material and losses occur.  These losses (at 
zero dc current) increase at a rate given by \Iac$^n$  where n is 
roughly 3.5, i.e. between the predictions of Norris for an elliptical 
cross section and a thin strip, a plausible result.  Thus a first step 
before attempting to fit the data to the theory discussed above would 
be to add a constant, equal to the appropriate plateau, to the losses 
calculated for each curve of Figure\ref{pertes_exp_plateau}.  However 
another hurdle appears when one notes the position of the end of the 
plateau say for the lowest curve with \Iac~=~4.2~A.  This excludes the 
model with \Isb since we are in the condition \Iac~$<$~\Isb/2, for 
which no extra losses are allowed.  Thus we are forced to exclude this 
model and to continue the analysis with Im instead.  A single fitting 
parameter of \Imm~=~21.8~A for all curves yields a relatively good fit 
of the data.  This fit is improved if, as in 
Figure\ref{pertes_exp_plateau}, \Ic is set at 27~A instead of the 
measured value of 29~A.

Higher values of ac current are considered in 
Figure\ref{pertes_exp_vallee}.  The data in the inset apply to the 
sample with \Ic~=~29~A.  A plateau and an increase in the losses is 
observed for \Iac~=~25.5~A but a minimum is finally observed for 
\Iac~=~31.1~A.  The fact that this is beyond Ic is not a problem in 
that the quoted value applies to the tape as a whole rather than only 
to the grains.  Also the criterion for determining \Ic is somewhat 
arbitrary.  A similar effect is observed with the critical current 
reduced to 25~A.  The first appearance of the minimum is observed with 
\Iac~=~26.2~A.  Due to the complex nature of these curves, no fit was 
attempted, the curves being merely a guide for the eye.  Nevertheless, 
the predicted minimum is approximately 50\% lower than the level 
without a dc current in agreement with the predictions shown in 
Figure\ref{pertes_im_vallee}.

As in Ref.\cite{leblanc96}, we consider either a Meissner current or a 
surface barrier current.  We are forced to reject the latter due to the 
position of the end of the a region defined by our experimental 
results.  On the other hand, as in Ref.\cite{ben98}, we cannot accept 
the Meissner current option without any questions.  Such a current 
would be of the order of $10^{-5}$ A whereas we obtain 21.8~A.  Again 
we must refer to the granular nature of the high temperature 
superconductor in the tape.  The flux lines penetrate easily into the 
intergrain material which has a low critical current.  An increase of 
the dc current forces the flux lines to penetrate into the grains 
thereby increasing the losses.  These even display two minima for 
\Iac~=~26.9~A, possibly a reflection of the excursion of flux lines in 
a cycle which include one and then two grains.  This picture will be 
elaborated in a forthcoming article \cite{ben99}.

\section*{\sc Conclusion}

These results add support to the observation by this group of a 
decrease of ac losses when a dc current is superimposed onto an ac 
current that is large with respect to a postulated \Imm or \Isb.  The 
detailed behavior is complicated by the granular nature of the 
silver-gold alloy sheathed Bi--2223 tape but the observed reduction of 
approximately 50\% of losses compared to the ones without a dc current 
correspond to the calculated result.  At lower ac current, computer 
simulations succeeded in fitting the data with a single parameter 
\Imm~=~21.8~A whereas a surface barrier model is incompatible with the 
data. Nevertheless the latter model should not be discarded completely 
since it considered the barrier as being at the surface of the tape as 
a whole although it is in reality at the surface of each grain.  
Finally we recall that for the first time, the simulations took into 
account the elliptical nature of the tape cross-section.

\subsection*{Acknowledgment}

This work was supported by Hydro-Québec and the Natural Sciences and 
Engineering Research Council of Canada


\begin{thebibliography}{12}

\bibitem{connel75} R.D. McConnell and P.R. Critchlow, {\em Rev. Sci. 
Instrum.}{\bf 46}:511 (1975).

\bibitem{bean64} Charles P. Bean, {\em  Review of Modern Physics} {\bf  
36}:31 (1964).

\bibitem{norris70} W.T. Norris, {\em J. Phys. D: Appl. Phys.} {\bf 
3}:489 (1970).

\bibitem{pat96} P. Dolez, M. Aubin, D. Willén, R. Nadi and J. Cave, 
{\em Supercond. Sci. Technol.} {\bf 9}:374 (1996) and references 
therein.

\bibitem{clem79} J.D. Thompson, M.P. Maley, and John R. Clem, {\em J. 
Appl. Phys.} {\bf 50}:3531 (1979).

\bibitem{ben98} P. Dolez, B. des Ligneris, M. Aubin, W. Zhu and J. 
Cave, {\em Proc. Int. Conf. Appl. Supercond., Palm Desert U.S.A.} Sept 
1998.  To be published.

\bibitem{oomen97} M.P. Oomen, J. Rieger, M. Leghissa and H.H.J. ten 
Kate, {\em Physica C} {\bf 290}:281 (1997).

\bibitem{leblanc96} M.A.R. LeBlanc,D.S.M. Cameron, D. LeBlanc and J. 
Meng,{\em  J. Appl. Phys.} {\bf  791}:334 (1996).

\bibitem{ben99} M. Aubin, B. des Ligneris, P. Dolez, and J. Cave, to be 
published.

\bibitem{pat98_2} P. Dolez, J. Cave, D. Willén, W. Zhu and M. Aubin, 
{\em Cryogenics} {\bf  38}:429(1998).

\end{thebibliography}
\end{document}